\documentclass{cimento}

\usepackage{graphicx}  

\title{A New Catalogue of Galactic Red Supergiants for direct detection of episodic mass-loss events}
\author{M.~Lauriano\from{ins:a}\from{ins:b}\ETC,
F.~Bocchino\from{ins:a},
M.~Miceli\from{ins:c}\from{ins:a},        
S.~Orlando\from{ins:a},
O.~Petruk\from{ins:a},
A.~Pastorello\from{ins:d},
M.~Limongi\from{ins:e}
    \atque
A.~Chieffi\from{ins:e}}

\instlist{\inst{ins:a} INAF-Osservatorio Astronomico di Palermo - Palermo, Italy
  \inst{ins:b} Dipartimento di Fisica e Astronomia, Universit\`a degli Studi di Padova - Padova, Italy
  \inst{ins:c} Dipartimento di Fisica e Chimica, Universit\`a degli Studi di Palermo - Palermo, Italy
  \inst{ins:d} INAF-Osservatorio Astronomico di Padova - Padova, Italy
  \inst{ins:e} INAF-Osservatorio Astronomico di Roma - Roma, Italy}

\begin{document}

\maketitle

\begin{abstract}
We present GalRSG, a new catalogue of 227 Galactic Red Supergiants in the Scutum–Crux region, which provides a nearly coeval and co-distant sample, target of a long-term, panchromatic, high-cadence photometric monitoring campaign aimed at detecting pre-supernova variability and luminous eruptively driven mass-loss events. 
\end{abstract}

\section{Introduction}

Red Supergiants (RSGs) represent one of the final evolutionary phases of massive stars with initial masses between $8$ and $25\, M_{\odot}$~\cite{ref:beasor}. During this phase, the star exhausts hydrogen in its core and evolves toward lower effective temperatures, typically in the range $T_{\mathrm{eff}} \approx (3300$–$4500$) K, on approximately constant luminosity of $\text{Log} \, (L_{\text{bol}}/L_{\odot}) \sim 4$–$5.8$, while its radius significantly increases~\cite{ref:levesque}. RSGs are regarded as prime candidates for core-collapse supernovae (CCSNe), particularly Type II-P events, thanks to the direct detections of RSG progenitors in pre-explosion imaging~\cite{ref:smartt09}.

Over the last decade, growing observational evidence has revealed that many CCSNe –particularly of Types IIP and IIn– are enshrouded in a dense and often inhomogeneous Circumstellar Medium (CSM), located in the close proximity to the progenitor star~\cite{ref:guarcello,ref:yaron}.
In parallel, current wide-field extragalactic transient surveys, which continuously monitor the sky for variable and explosive phenomena, have identified transients that are fainter than typical SNe but significantly more luminous than classical novae~\cite{ref:pastorello}. These events are interpreted as signatures of enhanced, episodic mass-loss in the final stages of massive-star evolution, leading to the formation of dense CSM around the progenitor~\cite{ref:fraser}. Mass-loss episodes range on timescales from thousands of years down to just a few days before core-collapse~\cite{ref:pastorello}. 
Such episodes in low-mass RSGs ($\lesssim 20\, M_{\odot}$) appear to be linked to the core activity during the late-stages of nuclear burning, and they are not included in the current stellar evolution models. A promising scenario is the wave-driven mass-loss, in which internal gravity waves propagating from the core to the stellar surface can be excited by vigorous convection bulks depositing energy that can unbind up to $\sim 1\, M_{\odot}$ of stellar material. Also nuclear flashes associated with the burning of heavy elements (O, Ne, Si) and binary interaction have been proposed as complementary causes of mass ejections~\cite{ref:leung,ref:fuller}.
Current extragalactic surveys are affected by a bias against the RSGs population with a luminosity $L \lesssim 10^{5}\,L_{\odot}$, and are further limited by the lack of homogeneous pre-SN observational data. Since an energy injection into the envelope is expected to produce detectable modifications in the progenitors' light curves, a systematic high-cadence monitoring programme of Galactic RSGs could uncover such photometric precursors and provide early-warning candidates for imminent SN event~\cite{ref:davies22}.

\section{The Catalogue}

The recent discovery of a number of nearly coeval and co-distant RSGs in the Scutum-Crux region has significantly increased the known Galactic population, enabling new studies of pre-SN outbursts. 
The Scutum-Crux region is located on the Galactic plane at longitude $l \sim 27^{\,\circ}$, where a starburst episode likely occurred $10$-$20\, \mathtt{Myr}$ ago as a consequence of the interaction between the Galactic bar and the Scutum spiral arm~\cite{ref:davies08}. From this region, we selected a sample of 227 RSGs to construct a new catalogue, named GalRSG, representing a nearly homogeneous population largely coeval and located at the comparable distance of $\sim(4-7)$ kpc. This uniformity in age and distance makes the sample ideal to investigate faint pre-SN activity in low-mass RSGs. The selected targets span the longitude range $10^{\circ} \lesssim l \lesssim 35^{\circ}$, with estimated masses of $\sim(8-15)\,M_\odot$ and luminosity $L \sim (10^4-10^{5.5})\, L_{\odot}$. The sample includes RSGs distributed both in clusters and in wide-field associations. The main Galactic associations considered in this work, and the number of objects included in each of them are shown in Table~\ref{tab:rsg_combined}.

\begin{table}[h!]
\caption{GalRSG catalogue.}
\label{tab:rsg_combined}
\centering
\renewcommand{\arraystretch}{0.9}
\begin{tabular}{lccc}
\hline
Association & Red Supergiants & Type & Reference \\
\hline
RSGC1                & 14      & Cluster    & \cite{ref:davies08} \\
Stephenson 2 (RSGC2) & 26      & Cluster    & \cite{ref:davies07} \\
RSGC3                & 15      & Cluster    & \cite{ref:clark} \\
Alicante 7           & 16   & Cluster    & \cite{ref:negueruela11} \\
Alicante 8           & 23   & Cluster    & \cite{ref:negueruela10} \\
Alicante 10          & 12   & Cluster    & \cite{ref:gonzalez} \\
\hline
Negueruela+12 & 60 & Wide-field & \cite{ref:negueruela12} \\
Messineo+16   & 61 & Wide-field & \cite{ref:messineo16} \\
\hline
\end{tabular}
\end{table}

\begin{figure}[h!]
    \centering
    \includegraphics[width=0.67\linewidth]{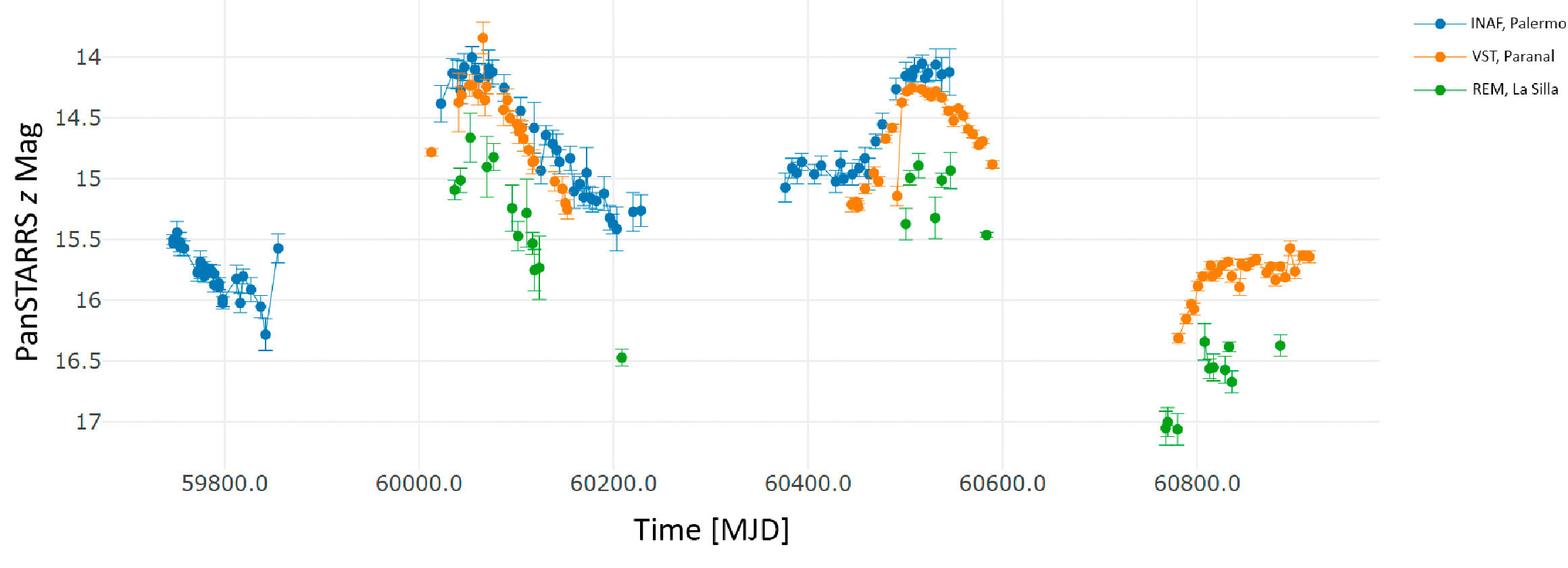}     
    \caption{Light curve in $z$ band of star \textit{2MASS J18443885-0326135}, a candidate RSG in Alicante 7. Each colour corresponds to one of the three facilities monitoring the target in the ongoing campaign. The REM data (green points) appear fainter, likely due to an overestimation of the background or a photometric calibration issue, which will be addressed in a forthcoming paper.}
    \label{fig:light-curve}
\end{figure}

In order to detect late-stage outbursts and compare them with recent models of the final phases of low-mass RSGs, we started a long-term, high-cadence, multi-band photometric monitoring campaign.
The complex spatial distribution of the stars listed in Table~\ref{tab:rsg_combined} required a tailored observational strategy for light curve acquisition. This strategy is based on an ongoing data-collection campaign that exploits the complementary strengths of multiple facilities: the VST at Paranal (Chile), with its wide-field of view, is used to survey field RSGs every $5 \pm2$ days in the Sloan $i$ and $z$ bands; REM at the La Silla observatory (Chile) sample the RSGs in clusters every $\sim 2$ days in the optical $g$,$r$,$i$ and the near-IR $J$,$H$,$K$ filters. Additionally, a pathfinder campaign has been carried out in 2022 using the 36 cm SC telescope at INAF–Palermo, still used in the $i$ and $z$ bands for the brightest subset of stars. An example of light curve is shown in Fig.~\ref{fig:light-curve}.

\begin{figure}[h!]
    \centering
    \includegraphics[width=0.63\linewidth]{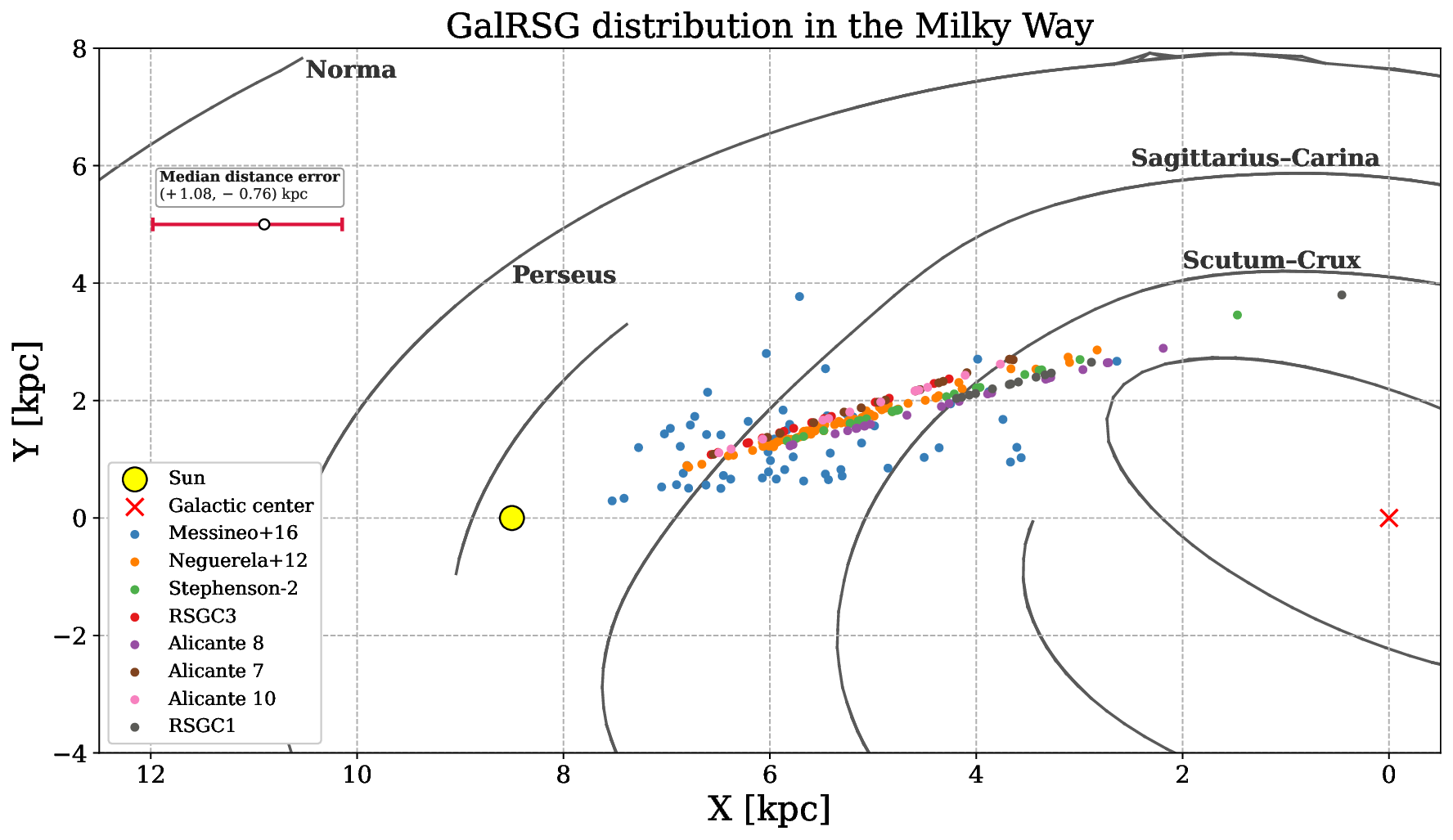}     
    \caption{Distribution of GalRSG stars in the Galactic plane, using the best photogeometric distances values from~\cite{ref:bailerjones}. The black curves trace the main spiral arms of the Milky Way, with the Sun (yellow circle) and the Galactic centre (red cross) shown for reference. The uncertainty is shown by the median value in the top-left part as a red bar.}
    \label{fig:galaxymaprsg}
\end{figure}

Reliable distance estimates are crucial to derive the physical parameters of the stars in the catalogue and the interstellar extinction, but are difficult to obtain in the highly obscured Galactic  region, such as Scutum. We therefore used Gaia EDR3 astrometry and adopted the photogeometric distances from~\cite{ref:bailerjones} to mitigate the poorly constrained parallaxes. As shown in Fig.~\ref{fig:galaxymaprsg}, these distances are affected by large uncertainties, typically of the order of $1$ kpc, reflecting the limited Gaia sensitivity in the near-IR, where RSGs emit most, and the strong interstellar extinction along the line of sight.

A major limitation in the current studies of Galactic RSGs is the paucity and heterogeneity of published stellar parameters. For most stars in the GalRSG sample, only partial information is available in the literature, often restricted to spectral classifications, while key quantities such as temperature, extinction, bolometric luminosity, radius, and mass-loss rates are missing or derived using assumptions. This fragmentation inhibits a uniform characterization of the population and hampers comparative studies across different environments and evolutionary stages.
In~\cite{ref:lauriano} we report the table of our GalRSG catalogue, along with the stellar parameters derived from a literature compilation.

\section{Conclusions}

We have presented GalRSG, a new catalogue of 227 Galactic RSGs in the Scutum--Crux region, providing a largely coeval and co-distant sample that is well suited for systematic studies of late-stage massive star evolution. By combining the homogeneous target selection with a long-term, high-cadence, multi-band photometric monitoring campaign, this work establishes a dedicated observational framework to search for low-amplitude variability and pre-SN photometric activity in Galactic RSGs.

We have derived literature-based stellar parameters for all our sources. However, the lack of homogeneous data, combined with the poorly constrained and systematically underestimated distances with respect to the Galactic bar–Scutum arm intersection region, require further analysis and a homogeneous determination of the physical properties of all targets based on a consistent methodology, to obtain a truly homogeneous sample. The ongoing 5-years monitoring campaign started in 2023 is still ongoing. The complete characterization of the stellar parameters for all the RSGs in our sample will be the subject of a forthcoming paper.

\acknowledgments
F.B. and M.L. acknowledge support of INAF grant GO-GTO2024 \textit{"Pre-supernova outbursts in Galactic RSGs"}. Based on data collected with the INAF VST telescope at the ESO Paranal Observatory and based on observations made with the REM Telescope, INAF Chile. F.B., M.M., S.O., and M.L. acknowledge financial contribution from the PRIN 2022 (20224MNC5A) \textit{"Life, death and after-death of massive stars"} funded by European Union - Next Generation EU. A.P. acknowledges support of the PRIN-INAF 2022 \textit{"Shedding light on the nature of gap transients: from the observations to the models"}.

\end{document}